\documentclass[twocolumn,showpacs,amsmath,amssymb,prl]{revtex4}
\usepackage{graphicx}% Include figure files
\usepackage{dcolumn}% Align table columns on decimal point
\usepackage{bm}% bold math

\begin{document}
\input{epsf}

\title{Cosmology with Hypervelocity Stars}

\author{Abraham Loeb}

\affiliation{Institute for Theory \& Computation, Harvard University,
60 Garden St., Cambridge, MA 02138, USA}

\begin{abstract} 

In the standard cosmological model, the merger remnant of the Milky
Way and Andromeda (Milkomeda) will be the only galaxy remaining within
our event horizon once the Universe has aged by another factor of ten,
$\sim 10^{11}$ years after the Big Bang.  After that time, the only
extragalactic sources of light in the observable cosmic volume will be
hypervelocity stars being ejected continuously from Milkomeda.
Spectroscopic detection of the velocity-distance relation or the
evolution in the Doppler shifts of these stars will allow a precise
measurement of the vacuum mass density as well as the local matter
distribution. Already in the near future, the next generation of large
telescopes will allow photometric detection of individual stars out to
the edge of the Local Group, and may target the $\sim 10^{5\pm 1}$
hypervelocity stars that originated in it as cosmological tracers.

\end{abstract}

\date{\today}
\maketitle

\paragraph*{Introduction.}
In the standard LCDM cosmological model \cite{Komatsu}, the vacuum
mass density which just started dominating the cosmic mass budget will
remain constant
\begin{equation}
\rho_{\rm v} = 0.7\times 10^{-29}~{\rm g~cm^{-3}}~,
\label{eq:vac}
\end{equation}
and so in the future, the scale factor of the Universe will grow
exponentially with time $a\propto \exp\{H_{\rm v}t\}$, at a rate
\begin{equation}
H_{\rm v}={\dot{a}\over a}=\left({8\pi G\rho_{\rm v}\over
  3}\right)^{1/2}= (1.6\times 10^{10}~{\rm yr})^{-1} .
\end{equation}

This accelerated expansion has important consequences for observers
once the Universe ages by merely 1--2 orders of magnitudes from now.
Within $\sim 10^{11}$ yr after the Big Bang, all galaxies outside the
Local Group will exit from our event horizon \cite{Loeb,Nagamine}.  At
that time, the Local Group itself will be a single galaxy, Milkomeda,
due to the imminent merger between the Milky Way and Andromeda in only
$\sim 5\times 10^9$ yr from now \cite{Cox}. Milkomeda will be
dominated by stars of low masses $m_\star\sim 0.1$--$1M_\odot$, for
which the lifetime $t_\star \approx 6.3\times
10^{12}(m_\star/0.1M_\odot)^{-2.8}~{\rm yr}$ \cite{Adams} is
comparable to the age of the Universe, $t$.  Within $\sim 67~e$--folds
or $\sim 10^{12}~{\rm yr}$, the wavelength of the cosmic microwave
background photons will be stretched by a factor of $\sim 10^{29}$ and
exceed the scale of the horizon $R_{\rm hor}\sim cH_{\rm
v}^{-1}=4.9\times 10^{3}~{\rm Mpc}$. At that time, not only external
galaxies but also the nearest extragalactic protons will be pushed out
of our horizon and not be available for tracing the cosmic expansion.
This realization has led to the naive expectation that an empirical
reconstruction of the past cosmic history will become impossible
\cite{Ellis,Krauss,AL,Island}. Will future observers be unable to
verify the validity of the standard cosmological model?

Here we show that the continuous flow of hypervelocity stars escaping
Milkomeda will in fact allow future observers to measure $\rho_{\rm
v}$ and the mass distribution within Milkomeda. This information will
be sufficient to demonstrate the validity of the standard cosmological
model at the above mentioned times.
 
\paragraph*{Cosmology with Hypervelocity Stars.}
Hypervelocity stars (HVSs) have velocities in excess of the escape
velocity of their host galaxy. Such stars were discovered in the Milky
Way halo over the past six years \cite{Brown} and are thought to be
produced by stellar interactions with the nuclear black hole SgrA*.
Although most of the known HVSs are bright and short-lived due to
observational selection effects, one expects that lower mass stars are
more commonly ejected from the Galactic center \cite{Gould}. The
ejection rate is low, once per $\sim 10^{5\pm 1}$ yr, and is therefore
expected to continue into the distant future long before the central
reservoir of stars will be depleted in the Milky Way or Milkomeda.

For a spherically-symmetric mass distribution, the orbit of an HVS is
purely radial, starting with some ejection velocity $\gtrsim 10^3~{\rm
km~s^{-1}}$ at $r\sim 0$ and subjected to two radial acceleration
terms \cite{Lahav} at $r\ll R_{\rm hor}$,
\begin{equation}
{d^2r\over dt^2}=-{GM(<r)\over r^2} + H_{\rm v}^2 r ,
\label{acc}
\end{equation}
with $M(<r)$ being the galaxy mass interior to a physical radius $r$.
For a total Milkomeda mass of $M_{\rm tot}= 10^{13}M_{13}~M_\odot$,
the second (``cosmological constant'') term will start to dominate at
radii beyond
\begin{equation}
R_{\rm tran}=\left({GM_{\rm tot}\over H_{\rm v}^2}\right)^{1/3} =2.3
M_{13}^{1/3}~{\rm Mpc} .
\label{tran}
\end{equation}
Figure 1 shows the relative contributions of the two acceleration
terms on the right-hand-side of equation (\ref{acc}). The dominance of
the second term at large distances motivated already the suggestion
that the vacuum-dominated state of the future Universe could be probed
with rocket flights \cite{Lewis}. Here we consider naturally produced
HVSs in place of artificially made rockets to retrieve this
information.

\begin{figure}[th]
\includegraphics[scale=0.4]{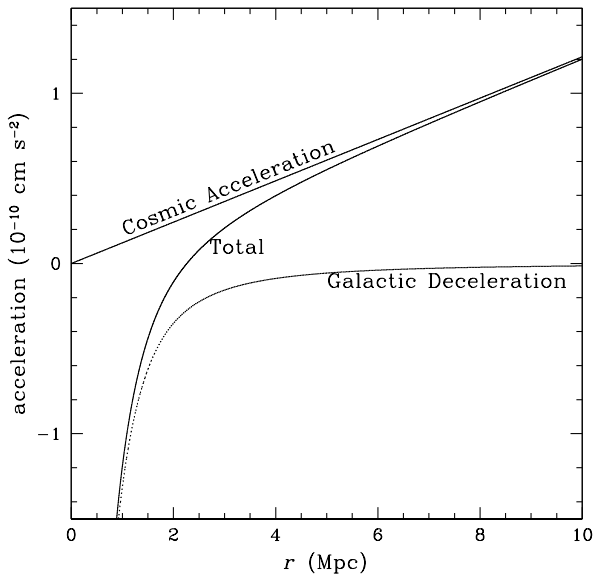}
\caption{The net acceleration of an HVS in Milkomeda as a function of
radius $r$ (middle line).  The Galactic deceleration (lower line) is
calculated for a halo with a total mass $M_{\rm tot}=10^{13}M_\odot$,
a virial radius $R_{\rm vir}=300~{\rm kpc}$, and a truncated density
profile described by Eq. (1) of Ref. \citep{Busha} (with
$R_{200}\equiv R_{\rm vir}$) that fits related numerical
simulations. Half of the halo mass lies outside $R_{\rm vir}$. The
acceleration due to the cosmological constant (upper line) dominates
over the Galactic deceleration at large radii, $r>R_{\rm tran}=
2.3~{\rm Mpc}$ (see Eq. \ref{tran}).  }
\end{figure}

With a typical escaping speed of $\sim 10^3~{\rm km~s^{-1}}$, an HVS
will traverse $R_{\rm tran}$ in a time $\sim 2\times 10^9$ yr, much
shorter than the lifetime of most of the stars ejected from
Milkomeda's center.  By using advanced technologies, future observers
could measure the acceleration of those stars from the change in the
Doppler shift of their spectral lines. The distance of those stars can
be found by inferring the intrinsic stellar luminosities from detailed
spectroscopic observations.  Measurements beyond $R_{\rm tran}$ will
determine the value of $\rho_{\rm v}$, whereas closer HVSs will be
used to find Milkomeda's mass distribution, $M(<r)$.

Since Milkomeda is surrounded by vacuum, it will have a sharp outer
boundary \cite{Busha}. The boundary is interior to $R_{\rm tran}$
because matter outside $R_{\rm tran}$ is accelerated outwards and
cannot be bound gravitationally. The average matter density within the
virial radius of Milkomeda $R_{\rm vir}= 300 R_{300}~{\rm kpc}$ can
then be measured,
\begin{equation}
\rho_{\rm vir}=3\times 10^{-27}~\left[{M(<R_{\rm vir})\over
5\times 10^{12}~M_\odot}\right] R_{300}^{-3}~{\rm g~cm^{-3}}.
\end{equation}
The spherical collapse model implies that a galaxy embedded in a
uniform matter-dominated background at early times should virialize
with a mean density of $\sim 100$--$200$ times the background matter
density, $\bar{\rho}_{\rm m}$ \cite{Gunn}. Future observers will then
be able to estimate that the mean background density when Milkomeda
had formed was $\bar{\rho}_{\rm m} \sim (3\times 10^{-27}~{\rm
g~cm^{-3}}/200)= 1.5\times 10^{-29}~{\rm g~cm^{-3}}$.  This value is
coincidentally comparable to the vacuum mass density in equation
(\ref{eq:vac}). The age of the Universe when Milkomeda had formed can
then be estimated to be $\sim H_{\rm v}^{-1}$, since the total Hubble
expansion rate $H=\dot{a}/a$ satisfies $H^2=(8\pi G/3)(\bar{\rho}_{\rm
m}+\rho_{\rm v})$. Because the matter density is diluted as
$\bar{\rho}_{\rm m}\propto a^{-3}$, one will then be able to conclude
that the Universe must have been dominated by matter at earlier times.

The elapsed time since Milkomeda assembled most of its gas would be
measurable from the age envelope of its stars and white dwarfs,
$t_\star$. Such a measurement will provide a good estimate for the
total time since the Big Bang, $t\approx (H_{\rm
v}^{-1}+t_\star)$. The average matter density at $t$ can then be
verified to be unmeasurably small, $\bar{\rho}_{\rm m}(t)\sim
\rho_{\rm v} \exp\{-3(H_{\rm v}t-1)\}$.

At $t\gtrsim 10^{12}~{\rm yr}$, the HVSs escaping from Milkomeda will
be the only extragalactic sources of light filling up the observable
volume of the Universe.  For stars ejected with an initial speed $v_0$
from Milkomeda's center at $r=0$, equation (\ref{acc}) can be
integrated to give the velocity $v(r)\equiv (dr/dt)$ as a function of
radius at distances $r\ll R_{\rm hor}$. Given a steady ejection rate
$\dot{N}$, the radial profile of the number density of stars $n(r)$
can be derived from the continuity equation,
\begin{equation}
n = {\dot{N}\over 4\pi r^2 v(r)} ~.
\label{continuity}
\end{equation}
At $r\sim R_{\rm tran}$ the velocity of an HVS is reduced to $v_{\rm
tran} \approx (v_0^2-v_{\rm esc}^2)^{1/2}$, where $v_{\rm esc}$ is the
escape speed from the production site of HVSs in the core of
Milkomeda.  Beyond $R_{\rm tran}$ the HVS velocity, $v\approx [v_{\rm
tran}^2+H_{\rm v}^2(r^2-R_{\rm tran}^2)]^{1/2}$, remains nearly
constant out to a radius $R_{\rm vel}\equiv (v_{\rm tran}/H_{\rm v})=
16.3 (v_{\rm tran}/10^3~{\rm km~s^{-1}})~{\rm Mpc}$, at which the
cosmic acceleration starts to increase its value significantly.
Therefore, at intermediate radii $R_{\rm tran}\lesssim r\ll R_{\rm
vel}$ equation (\ref{continuity}) yields a power-law density profile,
$n\propto r^{-2}$.  This profile extends down to $r=0$ if $v_0\gg
v_{\rm esc}$. In this regime, the total number of HVSs increases
linearly with distance. However, at much larger distances $R_{\rm vel}
\ll r\ll R_{\rm hor}$ the initial velocity is unimportant, and
$v\approx H_{\rm v} r$, allowing one to measure $\rho_{\rm v}$ from
the slope of the Hubble diagram ($v \propto r$) rather than from the
acceleration in equation (\ref{acc}).  In this remote region
$n\propto r^{-3}$, and the total number of HVSs grows only
logarithmically with distance, providing a limited statistical benefit
from improvements in the flux detection threshold of HVSs. Closer to
the horizon, the cosmological redshift (which is ignored here for
simplicity) makes the total number counts even flatter with decreasing
flux. For stellar masses $m_\star \gtrsim 0.5M_\odot$, the HVS counts
saturate at a sub-horizon distance for which the travel time equals
the HVS lifetime.  For example, solar-mass stars with a lifetime
$t_\star=10^{10}~{\rm yr}$ can only shine while they travel out to a
distance of $\sim 30$ Mpc for an ejection speed $v_0= 3\times
10^3~{\rm km~s^{-1}}$ and merely $\sim 7$ Mpc for $v_0=10^3~{\rm
km~s^{-1}}$.

Since the observed energy flux from a star of a given luminosity
$L_\star$ is $F_\star=L_\star/(4\pi r^2)$, the total number of HVSs in
the sky above an observed flux $F$ is given by
\begin{equation}
N(>F)= 4\pi \int_0^{r_{\rm max}(F)} n r^2 dr ,
\label{number} 
\end{equation}
where $r_{\rm max}(F)\equiv (L_\star/4\pi F)^{1/2}$. Figure 2 shows
the resulting flux distribution, normalized by an HVS ejection rate of
$\dot{N}=10^{-5}~{\rm yr^{-1}}$. For stars with $m_\star\gtrsim
0.5M_\odot$, this distribution should be trimmed to flatten at the
minimum flux corresponding to the distance for which the travel time
equals the HVS lifetime.  The travel time to a radius $r$ can be read
directly from Figure 2, since it equals $(N/\dot{N})=10^5 N~{\rm yr}$
in a steady state.  Stars with $m_\star =0.5M_\odot$ and $L_\star
=0.07L_\odot$ (assumed to be forming steadily out of the gas reservoir
in Milkomeda's nucleus) are the most luminous HVSs which could in
principle be observed out to the horizon distance $R_{\rm hor}$, since
their lifetime $t_\star\sim 7\times 10^{10}~{\rm yr}$ is comparable to
the travel time from Milkomeda to $\sim R_{\rm hor}$. A civilization
living around an M-dwarf HVS with $m_\star\lesssim 0.5 M_\odot$ will
have the privilege of taking a {\it habitable} one-way trip beyond the
horizon of Milkomeda. They might find analogs of Milkomeda beyond its
horizon, although they are most likely to find empty space since other
galaxies are also separated away from them at an accelerated rate.

\begin{figure}[th]
\includegraphics[scale=0.4]{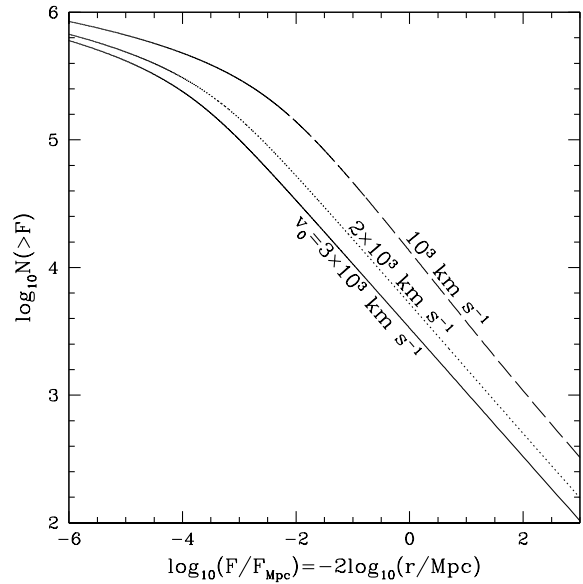}
\caption{Number of HVSs out to a distance $r$ (in Mpc) or with an
observed flux $>F$ (in unit of $F_{{\rm Mpc}}\equiv
L_\star/(4\pi\times {\rm Mpc}^2$).  Results are shown for an HVS
ejection rate $\dot{N}=10^{-5}~{\rm yr^{-1}}$ and the same mass
distribution of Milkomeda that was used in Fig. 1.  For long-lived
stars with $L_\star=0.07L_\odot$ ($m_\star=0.5M_\odot$ and
$t_\star=7\times 10^{10}~{\rm yr}$), the flux unit $F_{\rm Mpc}=
2.34\times 10^{-18}~{\rm erg~cm^{-2}~s^{-1}}$ corresponds to a
specific flux of $\sim 0.8~{\rm nJy}$ or an AB magnitude of 31.7 at a
wavelength of $\sim 1\mu$m and is comparable to the expected 1-$\sigma$
photometric sensitivity of the {\it James Webb Space Telescope (JWST)}
\cite{JWST}.  The different lines correspond to different ejection
speeds at $r=0$, namely $v_0= 10^3~{\rm km~s^{-1}}$ (upper line),
$2\times 10^3~{\rm km~s^{-1}}$ (middle line), and $3\times 10^3~{\rm
km~s^{-1}}$ (lower line).}
\end{figure}

\paragraph*{Conclusions.}
In the distant future, runaway HVSs from Milkomeda will be the only
extragalactic sources of light filling up the observable volume of the
Universe.  We have shown that a spectroscopic detection of the
evolution in their Doppler shifts could be used by future observers to
validate the standard cosmological model even at a time when the
wavelength of the relic radiation from the hot Big Bang exceeds the
scale of the horizon.

Indirect clues might also be available.  The existence of an early
radiation-dominated epoch could be inferred by measuring the abundance
of light elements in metal-poor stars and interpreting it with a
theory of Big Bang nucleosynthesis. The mass fraction of baryons
within Milkomeda could be assumed to be representative of the mean
cosmic value at early times. The nucleosynthesis theory can then be
used to find the necessary radiation temperature $T_\gamma\propto
a^{-1}$, such that the correct light element abundances would be
produced.  This would lead to an estimate of the time when matter and
radiation had the same energy densities. Since density perturbations
grew mainly after that time, it will be possible to estimate the
amplitude of the initial density fluctuation on the mass scale of
$M_{\rm tot}$ that was required for making Milkomeda at a time
$(t-t_\star)\sim H_{\rm v}^{-1}$ after the Big Bang. Without a
radiation-dominated epoch, this amplitude could have been arbitrarily
low at arbitrarily early times.  Future astronomers may already have
cosmology texts available to them, but even if they do not, we have
outlined a methodology by which they will be able to arrive at, and
empirically verify, the standard cosmological model.

The working assumption so far was that the vacuum energy density
remains constant and does not decay to a lower energy state over
hundreds of $e$-folding times. Measuring the dynamics of hypervelocity
stars at late cosmic times provides the added benefit of testing
whether the cosmological constant is truely constant. Indications to
the contrary might reveal new physics, such as that considered by
evolving dark energy models \cite{evolve}.

One might wonder whether the signal shown in Figure 1 can be used
today to measure $\rho_{\rm v}$ outside the Local Group.
Conceptually, such a measurement would be contaminated by the large
matter inhomogeneities that exist at distances smaller than tens of
Mpc around the Virgo cluster. But even if one were interested in
mapping the mass distribution of those inhomogeneities \cite{Sherwin},
the instrumental requirements are extremely challenging. With
state-of-the-art spectrographs using a laser frequency comb
\cite{Science} installed in the next generation of large telescopes
\cite{ELT}, one might hope to achieve at best a velocity precision of
$\sim 1~{\rm cm~s^{-1}}$, which corresponds to an acceleration
sensitivity of $\sim 3\times 10^{-9}~{\rm cm~s^{-2}}$ over a period of
a decade. Unfortunately, this sensitivity threshold is still much
larger than the expected acceleration amplitude of $\sim 10^{-10}~{\rm
cm~s^{-2}}$ at the largest possible distance of $\sim 14~{\rm Mpc}$,
which an HVS with a velocity of $\sim 10^3~{\rm km~s^{-1}}$ could
reach during the current age of the Universe. Although one could
attempt to measure the velocity drift over a longer period of time,
the actual acceleration sensitivity will be much worse than the
best-case value stated above given the faintness of a single star at
that distance (see Fig. 2).  However, measuring the distance and
velocity rather than acceleration of HVSs within the Local Group
should already become feasible within the next decade.  Future
generations of experimentalists will have $\sim 10^{12}~{\rm yr}$ to
improve upon this instrumental performance before it becomes crucial
for validating the standard cosmological model.

The quantitative results presented in Figures 1 and 2 may inspire
related observational work in the near future. In particular, a $\sim
2M_\odot$ star would be detectable photometrically by JWST (with a
signal-to-noise ratio $\gtrsim 10$) out to distances as large as a few
Mpc.  Thus, the next generation of large telescopes will allow
photometric detection of individual stars beyond the edge of the Local
Group and might target the $\sim 10^{5\pm 1}$ hypervelocity stars that
originated in it as cosmological tracers.

\bigskip
\bigskip
\paragraph*{Acknowledgments.}
I thank F. Dyson, D. Maoz, R. O'Leary, J. Pritchard, G. White, and an
anonymous referee for helpful comments.  These brief notes were
written in response to questions from the audience at a recent public
lecture that I gave about the future of our Universe. The work was
supported in part by NSF grant AST-0907890 and NASA grants NNX08AL43G
and NNA09DB30A.

\end{document}